\documentclass{article}
\usepackage{
    spconf,
    amsmath,
    graphicx,
    xcolor,
    amsfonts,
    booktabs,
    siunitx,
    tablefootnote,
    tikz
}
\usepackage[hidelinks]{hyperref}
\usepackage[font=small,labelfont=bf]{caption}

\title{CONDITIONAL DIFFUSION MODEL FOR TARGET SPEAKER EXTRACTION}
\name{Theodor Nguyen{\normalfont\textsuperscript{1}}, Guangzhi Sun{\normalfont\textsuperscript{1}}, Xianrui Zheng{\normalfont\textsuperscript{1}}, Chao Zhang{\normalfont\textsuperscript{2}}, Philip C Woodland{\normalfont\textsuperscript{1}}\thanks{This work has been performed using resources provided by the Cambridge Tier-2 system operated by the University of Cambridge Research Computing Service funded by EPSRC Tier-2 capital grant EP/T022159/1. Theodor Nguyen is supported by the Studienstiftung des Deutschen Volkes. The code will be made available after the review process.}}
\address{{\normalfont\textsuperscript{1}}Department of Engineering, University of Cambridge, Cambridge, UK
\\
{\normalfont\textsuperscript{2}}Department of Electrical Engineering, Tsinghua University, Beijing, China}

\begin{document}
\ninept
\maketitle
\begin{abstract}
We propose DiffSpEx, a generative target speaker extraction method based on score-based generative modelling through stochastic differential equations.
DiffSpEx deploys a continuous-time stochastic diffusion process in the complex short-time Fourier transform domain, starting from the target speaker source and converging to a Gaussian distribution centred on the mixture of sources.
For the reverse-time process, a parametrised score function is conditioned on a target speaker embedding
to extract the target speaker from the mixture of sources.
We utilise ECAPA-TDNN target speaker embeddings and condition the score function alternately on the SDE time embedding and the target speaker embedding.
The potential of DiffSpEx is demonstrated with the WSJ0-2mix dataset, achieving an SI-SDR of 12.9~dB and a NISQA score of 3.56.
Moreover, we show that fine-tuning a pre-trained DiffSpEx model to a specific speaker further improves performance, enabling personalisation in target speaker extraction.
\end{abstract}
\begin{keywords}
Target speaker extraction, diffusion, score-based generative modelling, ECAPA-TDNN, personalisation
\end{keywords}

\section{Introduction}
Target speaker extraction (TSE) involves identifying and isolating the speech signal of a specified target speaker from a mixture of signals.
TSE systems find practical use in the context of personalised speech processing, where the user of a personalised system can set up a downstream task to only process their voice, e.g. with personalised voice assistants or mobile phone applications.
Spatial information (e.g. from a microphone array \cite{Flanagan-1985}),  audio-visual cues \cite{Hershey-2001, Ochiai-2019, Sato-2021}) and auditive cues (e.g. fixed-length speaker embeddings \cite{Zmolikova-2019, Liu-2023, Xu-2020, Ge-2020, Wang-2021}) can be used to target and extract the correct speaker.
The presented work only considers auditive cues in the form of fixed-length speaker embeddings as conditioning vectors in the TSE task.
Such fixed-length embeddings are either provided by pre-trained speaker embedding models \cite{Zmolikova-2019, Liu-2023} or jointly learned in a multi-task training objective \cite{Xu-2020, Ge-2020, Wang-2021}. All of the above-mentioned TSE models are based on discriminative methods trained on signal similarity to the target signal by performing signal masking.
\begin{figure}[t]
  \includegraphics[width=8.5cm]{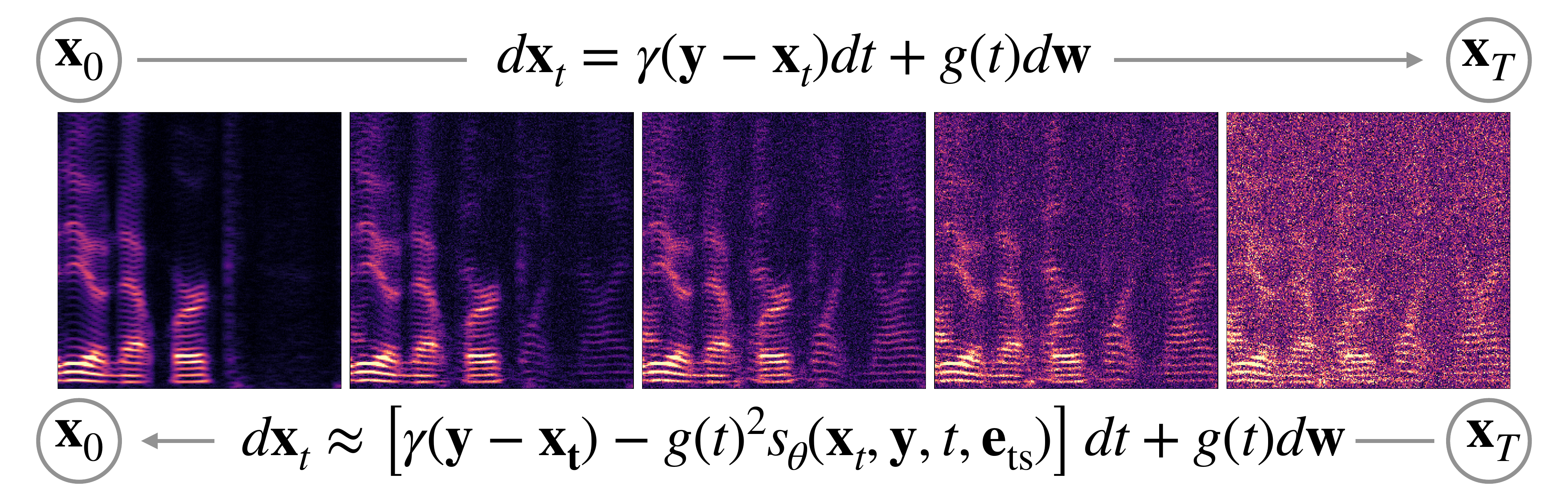}
\small\caption{The stochastic diffusion process (f.l.t.r) progressively perturbs the target speaker signal with Gaussian noise while interpolating from the target signal to the mixture. The reverse process (f.r.t.l) conditioned on the target speaker embedding $\mathbf{e}_\mathrm{ts}$ extracts the target speaker signal from the mixture of signals.}
\label{fig:sde-rsde}
\end{figure}

In contrast, generative methods aim to fit a parametrised model to the underlying data distribution. Diffusion-based generative models have recently shown state-of-the-art capabilities in image generation \cite{Ho-2020, Song-2020-2, Karras-2022, Song-2020-3, Song-2023} and speech processing \cite{Kong-2020, Lu-2021, Lu-2022, Welker-2022, Richter-2023, Lay-2023, Lemercier-2023}.
Denoising diffusion probabilistic models (DDPM, \cite{Sohl-2015, Ho-2020}) and score-based generative modelling through stochastic differential equations (SGM, \cite{Song-2020-2}) form the two main frameworks in diffusion-based modelling.
For the scope of this work, SGM is considered the framework of choice, although \cite{Song-2020-2} and \cite{Karras-2022} showed that DDPMs and SGMs are equivalent formulations.
SGMs have been applied to speech enhancement \cite{Welker-2022, Richter-2023, Lay-2023, Lemercier-2023} and blind source separation (BBS) for speech data \cite{Scheibler-2023}.

However, as far as we are aware, an SGM method for TSE is yet to be investigated. In the proposed DiffSpEx method (Section \ref{sec:method}), TSE is performed by running the reverse-time SDE in the complex short-time Fourier transform (STFT) domain, starting from a mixture of signals with a trainable \textit{score function} conditioned on fixed-length target speaker embeddings from the ECAPA-TDNN speech extractor \cite{Desplanques-2020}.
The proposed method is illustrated in Fig. \ref{fig:sde-rsde}.
The potential of DiffSpEx is demonstrated in Section \ref{sec:experiments} on a \textit{general} TSE task and a \textit{personalised} TSE task on the WSJ0-2mix dataset \cite{Baker-1992}, where it is shown in Section \ref{sec:results} that DiffSpEx has comparable performance to discriminative TSE methods in signal reconstruction (in SI-SDR \cite{LeRoux-2015}) while outperfoming them in uninstrusive speech quality and naturalness measured in NISQA \cite{Mittag-2021}. Moreover, it is shown that DiffSpEx can be fine-tuned to a specific speaker to improve signal reconstruction and speech quality further. Hence, the DiffSpEx method can be applied to settings where a particular (personalised) speaker is to be extracted with high quality and naturalness.
\section{Background}
In SGM, a data sample from the target distribution is progressively perturbed with Gaussian noise in a stochastic diffusion process to increase the support of the data distribution on the high-dimensional manifold.
This forward process is defined through a stochastic differential equation (SDE) \cite{Song-2020-2}.
The work in \cite{Anderson-1982} showed that the backward process to sample a target data point from (Gaussian) noise can be defined with a reverse-time SDE, where the required \textit{score function} can be trained as a deep neural network via score-matching \cite{Song-2020-2}.
\begin{tikzpicture}[remember picture,overlay]
    \node[draw=black, ultra thin, fill=white, text=black, anchor=south, 
    text width=\paperwidth-40mm,
    minimum height=1cm,
    inner sep=1mm,
    outer sep=0mm
    ] at ([yshift=6mm]current page.south) {
    \footnotesize © 2023 IEEE. Personal use of this material is permitted. Permission from IEEE must be obtained for all other uses, in any current or future media, including reprinting/republishing this material for advertising or promotional purposes, creating new collective works, for resale or redistribution to servers or lists, or reuse of any copyrighted component of this work in other works.
    };
\end{tikzpicture}
\label{sec:background}
\subsection{Score-based generative modelling through SDEs}
A stochastic diffusion process is a solution to an SDE and can be represented as a continuous forward process $\{\mathbf{x}(t)\}_{t\in[0,T]}$ starting at $t=0$.
The continuous forward process in SGM $\{\mathbf{x}(t)\}_{t\in[0,T]}$ can be formalised through the SDE \cite{Song-2020-2} as
\begin{equation}
\label{eq:SDE}
d\mathbf{x}=f(\mathbf{x}, t)dt+g(t)d\mathbf{w},
\end{equation}
where $f(\mathbf{x},t):(\mathbb{R}^{\text{d}_\mathbf{x}},\mathbb{R})\rightarrow\mathbb{R}^{\text{d}_\mathbf{x}}$
is called the drift coefficient and $\text{d}_\mathbf{x}$
denotes the dimensionality of $\mathbf{x}$, 
$g(t):\mathbb{R}\rightarrow\mathbb{R}$ is called the diffusion coefficient, and 
$\mathbf{w}\in\mathbb{R}^{\text{d}_\mathbf{x}}$ represents standard Brownian motion.
From a macroscopic view, the probability density function over all possible continuous stochastic trajectories $\{\mathbf{x}(t)\}_{t\in[0,T]}$ at a continuous timestep $t$ is denoted by $p_t(\mathbf{x})$.
For $t=0$, $p_0(\mathbf{x})$ is the data distribution $p(\mathbf{x})$, as there is no added noise at timestep $t=0$.
As the diffusion process proceeds to $t\rightarrow T$, i.e. the largest possible noise in the defined process, by design, $p_T(\mathbf{x})$ converges to a tractable distribution $\pi(\mathbf{x})$ called the \textit{prior distribution}.
With the SDE formulation of the diffusion process, data points can be sampled from the prior distribution by reversing the diffusion process.
In \cite{Anderson-1982}, it is proven that for every SDE of the form in Equation \eqref{eq:SDE}, there exists a corresponding reverse-time SDE
\begin{equation}
\label{eq:RSDE}
    d\mathbf{x}=\left[f(\mathbf{x}, t)-g(t)^2 \nabla_\mathbf{\mathbf{x}} \log p_t(\mathbf{x})\right] d t+g(t) d\mathbf{w},
\end{equation}
where $dt$ denotes a negative timestep in the reverse process and $\nabla_\mathbf{x} \log p_t(\mathbf{x})$ can be approximated by a trainable \textit{score function} $s_\theta(\mathbf{x}, t)$:
\begin{equation}
\label{eq:RSDE-score}
    d\mathbf{x}\approx\left[f(\mathbf{x}, t)-g(t)^2 s_\theta(\mathbf{x}, t)\right] d t+g(t) d\mathbf{w}.
\end{equation}
To sample a data point $\hat{\mathbf{x}}$ from $p(\mathbf{x})$, the reverse-time SDE in Equation \ref{eq:RSDE} can be solved with any numerical SDE solver by (i) sampling from the tractable prior distribution $p_T(\mathbf{x})=\pi(\mathbf{x})$ as a starting point and (ii) using the trained score model $s_\theta(\mathbf{x}, t)$ to get the correct reverse-time SDE.

\subsection{Speech enhancement in the complex STFT domain}
A diffusion-based speech enhancement model that operates in the complex STFT domain was proposed in \cite{Welker-2022} and \cite{Richter-2023}. The complex STFT spectrograms are transformed and retransformed with
\begin{equation}
\label{eq:STFT-transform}
    \tilde{c}=\beta|c|^\alpha \operatorname{e}^{i\angle(c)}\;\;\mathrm{and}\;\;
    c=\frac{1}{\beta}|\tilde{c}|^\frac{1}{\alpha }\operatorname{e}^{i\angle(\tilde{c})},
\end{equation}
respectively, where $c$ are the coefficients in the complex STFT, $\angle(\cdot)$ is the angle of the coefficients, $\alpha\in\mathbb{R}_+$ is a parameter that compresses the amplitude to upscale frequencies with lower energies, and $\beta\in\mathbb{R}_+$ is a scaling factor to scale the amplitudes to be between 0 and 1 for predictable noising in the diffusion-based model.
In their work, \cite{Welker-2022} and \cite{Richter-2023} proposed a speech enhancement method based on the SDE formulation of score-based generative models \cite{Song-2020-2} with the SDE
\begin{equation}
\label{eq:OUVE}
    d \mathbf{x}_t =\gamma(\mathbf{y}-\mathbf{x}_t) d t+g(t) d \mathbf{w},
\end{equation}
where $\mathbf{y}$ is the spectrogram of the noisy signal and $g(t)$ is the diffusion coefficient that is parametrised by $\sigma_{\min}$ and $\sigma_{\max}$ in
\begin{equation}
\label{eq:VE}
    g(t) =\sigma_{\min}\left(\frac{\sigma_{\max}}{\sigma_{\min }}\right)^t \sqrt{2 \log \left(\frac{\sigma_{\max}}{\sigma_{\min }}\right)}.
\end{equation}
The drift term $\gamma(\mathbf{y}-\mathbf{x}_t)$ reverts stochastic processes $\{\mathbf{x}_t\}_{t\geq0}$ that deviate from $\mathbf{y}$ back to $\mathbf{y}$, where $\gamma$ can be interpreted as the stiffness coefficient of the reversion effect.
Such a mean-reverting SDE is also called an Ornstein-Uhlenbeck SDE \cite{Ornstein-1930}.
Based on the form of the variance schedule $g(t)$, the SDE in Equation \eqref{eq:OUVE} is referred to as a \textit{Variance Exploding} SDE in \cite{Song-2020-2}, as the variance tends to infinity for $t \rightarrow \infty$.
In combination, \cite{Welker-2022} and \cite{Richter-2023} coined the SDE the Ornstein-Uhlenbeck Variance Exploding SDE (OUVE).
The closed-form solutions for the mean $\mu\left(\mathbf{x}_0, \mathbf{y}, t\right)$ and variance $\sigma(t)^2 \mathbf{I}$ for a Gaussian process are derived in Equations (5.50) and (5.53) in \cite{Sarkka-2019}:
\begin{equation}
\begin{aligned}
\label{eq:closed-form-mu}
\mu\left(\mathbf{x}_0, \mathbf{y}, t\right) =e^{-\gamma t} \mathbf{x}_0+\left(1-e^{-\gamma t}\right) \mathbf{y},
\end{aligned}
\end{equation}
\begin{equation}
\begin{aligned}
\label{eq:closed-form-sigma}
\sigma(t)^2 =
\frac{
    \sigma_{\min }^2\left(\left(\frac{\sigma_{\max }}{\sigma_{\min }}\right)^{2 t}-e^{-2 \gamma t}\right) \log \left(\frac{\sigma_{\max }}{\sigma_{\min }}\right)
}
{
    \gamma+\log \left(\frac{\sigma_{\max }}{\sigma_{\min }}\right)
}.
\end{aligned}
\end{equation}
\section{Method}
\label{sec:method}
The approach to performing TSE with a diffusion model is motivated by the promising performance of diffusion models in the BSS task \cite{Scheibler-2023}.
In the conclusion of \cite{Scheibler-2023}, the authors mentioned a gap in performance with state-of-the-art discriminative methods.
The authors argue that as both outputs come from the distribution of human speech, it is unclear to the generative model which parts of the mixture of human speech belong to which speaker.
In the experiments in Section \ref{sec:experiments}, we utilise fixed-length speaker embeddings to condition the diffusion-based generative model on a specific speaker distribution of the \textit{marginal} distribution of human speech to separate the target speaker from the mixture of speakers.

\subsection{Complex STFT features}
Denote the target speaker signal with $\mathbf{x}_0$ and the mixture signal with $\mathbf{y}$ in the complex STFT domain.
The fixed-length target speaker embedding vector is denoted by $\mathbf{e}_{\text{ts}}$.
First, $\mathbf{x}_0$ and $\mathbf{y}$ are transformed by the rescaling transformation from Equation \eqref{eq:STFT-transform}.
Both complex spectrograms $\{\mathbf{x}_0, \mathbf{y}\}\in\mathbb{C}^{B\times F}$, where $B$ is the number of frequency bins and $F$ is the number of time frames of the STFT, are passed as real and imaginary channels to the diffusion-based model, $\{\mathbf{x}_0, \mathbf{y}\}\in\mathbb{R}^{2\times B\times F}$.

\subsection{SDE formulation for target speaker extraction}
The SDE from Equation \eqref{eq:SDE} is used to continuously perturb the transformed complex STFT spectrogram $\mathbf{x}_0$ while the mean continuously shifts from $\mathbf{x}_0$ to $\mathbf{y}$ with Equation \eqref{eq:closed-form-mu}.
In practice, $\mathbf{x}_t$ is sampled with 
\begin{equation}
\begin{aligned}
\label{eq:x_t}
& \mathbf{x}_t=\mu\left(\mathbf{x}_0, \mathbf{y}, t\right)+\sigma(t)\,\mathbf{z},
\end{aligned}
\end{equation}
where $\sigma(t)$ is defined by Equation \eqref{eq:closed-form-sigma} and $\mathbf{z}$ is standard Gaussian noise.
The reverse-time SDE is defined by
\begin{equation}
\label{eq:RSDE-score-tse}
    d\mathbf{x}_t\approx
    \left[
        \gamma(\mathbf{y}-\mathbf{x_t})-g(t)^2 s_\theta(\mathbf{x}_t,\mathbf{y}, t, \mathbf{e}_{\text{ts}})
    \right] dt+g(t)d\mathbf{w},
\end{equation}
where $s_\theta(\cdot)$ is conditioned on the target speaker embedding $\mathbf{e}_{\text{ts}}$. 
Following the solution for $\nabla_{\mathbf{x}_t}\log p_{0t}\left(\mathbf{x}_t \mid \mathbf{x}_0, \mathbf{y}\right)$ of a Gaussian kernel $p_{0t}\left(\mathbf{x}_t \mid \mathbf{x}_0, \mathbf{y}\right)$ from \cite{Vincent-2011}, the loss function takes the form
\begin{equation}
\begin{aligned}
\label{eq:training-objective}
& \underset{\theta}{\arg \min } \:\mathbb{E}_{t, \mathbf{y}, \mathbf{z}, \mathbf{x}_t}
\left[
    \left\|\mathbf{s}_\theta\left(\mathbf{x}_t, \mathbf{y}, t, \mathbf{e}_{\text{ts}}\right)+\frac{\mathbf{z}}{\sigma(t)}\right\|_2^2
\right]
\end{aligned}
\end{equation}
as in \cite{Richter-2023}.
\section{Experimental Setup}
Two distinct DiffSpEx approaches are investigated: (i) a \textit{general} conditional TSE model (G-DiffSpEx), where the set of potential speakers is not limited and conditioned by the speaker embedding, and (ii) a \textit{personalised} TSE model (P-DiffSpEx), where a pre-trained model is fine-tuned on a specific speaker.
In the latter experiment, various total durations of reference signals $\{\SI{60}{\second}, \SI{180}{\second}, \SI{300}{\second}, \SI{600}{\second} \}$ are used to reveal how much reference audio is needed to fine-tune DiffSpEx on one speaker.
\label{sec:experiments}
\subsection{Datasets and speaker embedding}
The WSJ0-2mix \textit{min}-cut
\cite{Baker-1992} is chosen at a sample rate of 8~kHz to evaluate our proposed TSE model.
The datasets consist of 20,000 training (\SI{30}{\hour}), 5,000 validation (\SI{8}{\hour}) and 3,000 test utterances (\SI{5}{\hour}), where the set of speakers in the test dataset with 18 speakers is disjoint from the training and validation dataset (101 speakers).

\textbf{G-DiffSpEx}. The first speaker is set in the mixture signals to be the target speaker, and fixed-length target speaker embeddings are generated with ECAPA-TDNN \cite{Desplanques-2020} from a random sample of the clean signals of the respective speakers.
Additionally, random samples of one speaker are joined to at least \SI{60}{\second} signal length to generate target speaker embeddings with more context for evaluation, which we mark with \textit{60s}.

\textbf{P-DiffSpEx}. The individual speaker 40f from the WSJ0-2mix dataset is chosen to create proprietary \textit{personalised} datasets for different amounts of reference audio \{\SI{60}{\second}, \SI{180}{\second}, \SI{300}{\second}, \SI{600}{\second}\}, where in the personalised datasets, the target speaker audio is a randomly selected reference audio of the individual speaker 40f. The set of samples in the personalised training, validation and test set are disjoint, where the total duration of the set of samples for validation and testing is \SI{195}{\second}. Additionally, to report more representative results, the experiment is repeated for \SI{600}{\second} of reference audio with 10 speakers \{017, 01a, 01i, 01t, 01z, 029, 02e, 40f,  40l, 40o\}.

\subsection{Metrics}
The commonly used signal restoration metrics SI-SDR and SI-SDRi \cite{LeRoux-2015} in dB are used to evaluate the signal similarity between the extracted speaker signal and the ground truth signal in the time domain.
Furthermore, the intrusive speech quality metric PESQ \cite{Rix-2001} and the unintrusive
speech naturalness
metric NISQA \cite{Mittag-2021} are reported.

\subsection{Model and speaker embedding conditioning}
\begin{figure}[t]
\centering
  \includegraphics[width=8cm]{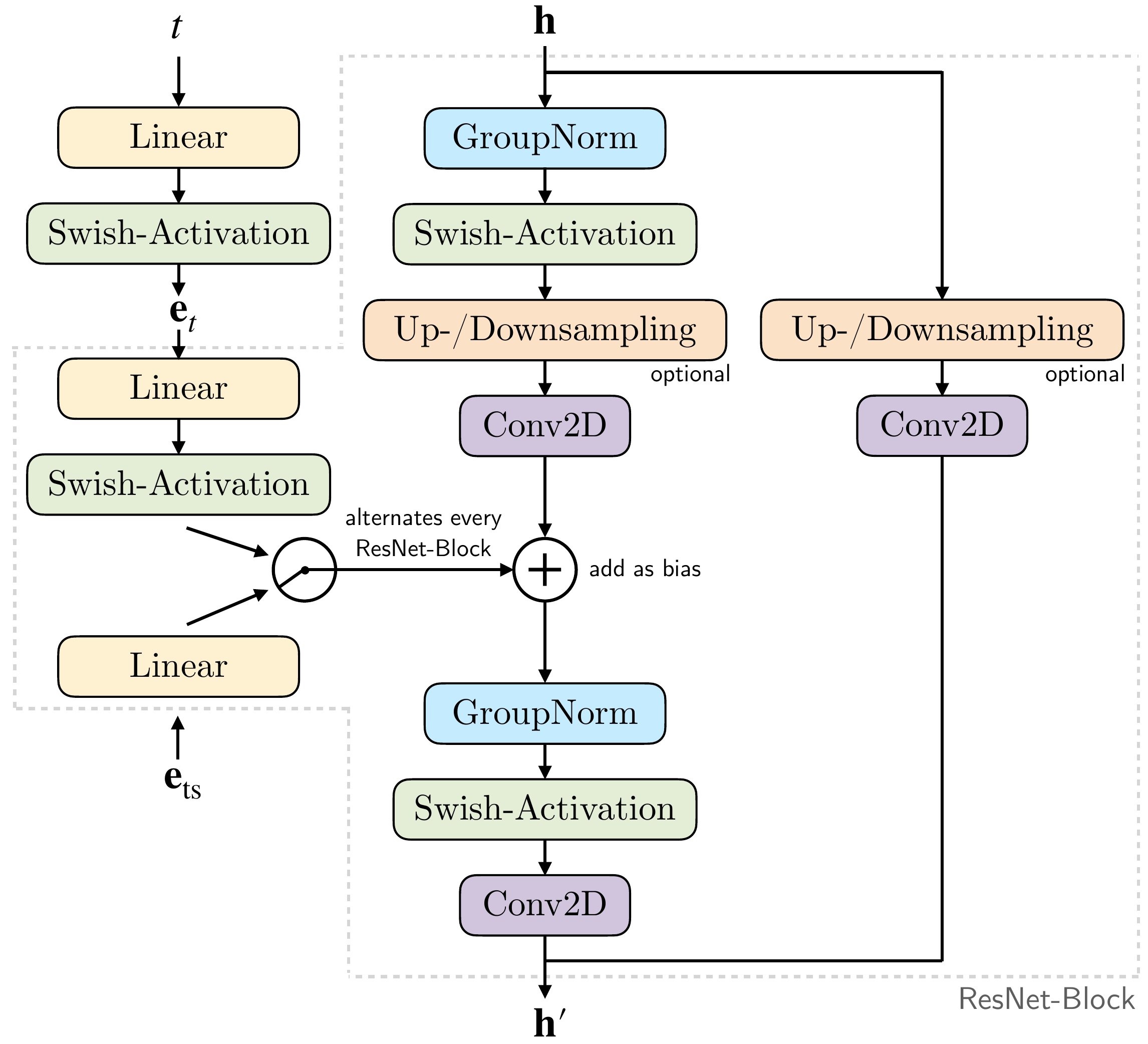}
\small\caption{ResNet-Block in the U-Net architecture with optional finite impulse response (FIR) filters for up- and downsampling.}
\label{fig:ResNet-Block}
\end{figure}

The parametrised score-function $\mathbf{s}_\theta(\cdot)$ is based on the noise-conditio\-ned score-matching network (NCSN++) from \cite{Song-2020-2}, a multi-channel U-Net neural network with residual connections \cite{Ronneberger-2015}.
As $\mathbf{s}_\theta(\cdot)$ is conditioned on the mixture signal $\mathbf{y}$ besides the sampled signal $\mathbf{x}_t$ \cite{Welker-2022, Richter-2023, Scheibler-2023}, the input to the NCSN++ score-function $(\mathbf{x}_t, \mathbf{y})\in\mathbb{R}^{4\times B\times F}$ has 4 channels \cite{Richter-2023}.
We use a smaller network configuration than \cite{Richter-2023, Scheibler-2023} of 4 encoder-decoder layers with 2 ResNet blocks each and no attention mechanism, resulting in a model with 37.7M parameters.
The ECAPA-TDNN \cite{Desplanques-2020} speaker embeddings $\mathbf{e}_{\text{ts}}\in\mathbb{R}^{192}$ and the time embeddings $\mathbf{e}_{t}\in\mathbb{R}^{512}$ from the variance scheduler $\sigma(t)^2$ are downsampled by feed-forward layers and alternately added as bias to the intermediate representations in the ResNet blocks. 
Fig. \ref{fig:ResNet-Block} illustrates such a ResNet block with alternated adding of the embeddings.

\subsection{Hyperparameters for inference and training}
The 8~kHz audio signals are transformed into the complex STFT domain with $N_{\mathrm{FFT}}=254$, window size = 256, hop length = 64 and $F=256$, resulting in $B=128$ and model input frames of 2048~ms length at \SI{8}{\kilo\hertz}.
The parameters for the complex STFT transformations from Equation \eqref{eq:STFT-transform} are set to $\alpha=0.5$ and $\beta=0.15$ as in \cite{Welker-2022} and \cite{Richter-2023}.
The parametrisation for the SDE in Equations \eqref{eq:RSDE-score-tse} and \eqref{eq:VE} are $\gamma=2.0$, $\sigma_{\min}=0.05$ and $\sigma_{\max}=0.5$.
For inference, a predictor-corrector sampler \cite{Song-2020-2} with the Euler-Maruyama solver and annealed Langevin dynamics \cite{Pastor-1994} is used, where the sampling parameters are set to $N=30$ and $r=0.5$.
The models are evaluated on SI-SDR of 100 random validation samples to choose the best checkpoint for test evaluation in order to save computational costs.

\textbf{G-DiffSpEx}. The G-DiffSpEx model is trained for 1000 epochs with an effective batch size of 48 on one NVIDIA-A100 GPU, which requires 250~h. We train the model with the Adam optimiser \cite{Kingma-2014}, a learning rate \num{5e-4} and a linear warm-up schedule of 2000 training steps.

\textbf{P-DiffSpEx}. The P-DiffSpEx model is fine-tuned on the pre-trained G-DiffSpEx model for 15 epochs with an effective batch size of 48 on one NVIDIA-A100 GPU, which requires 4~h. We fine-tune the model with the Adam optimiser at a learning rate of \num{5e-4}.
\section{Results}
\label{sec:results}
\subsection{General target speaker extraction}
The evaluation results on the trained DiffSpEx model for randomly chosen speaker embeddings and \textit{60s}-embeddings are given in Table \ref{tab:TSE-results}, where numbers with reported standard deviations (after $\scriptstyle\pm$) are evaluated on the same test script and numbers without standard deviations are taken from their respective publications.
The proposed method achieves an SI-SDRi better than earlier discriminative TSE models with better NISQA values. However, there is a noticeable gap in SI-SDR to the state-of-the-art discriminative TSE methods.
The NISQA score of 3.56 indicates that the extracted speech signals of DiffSpEx are more natural than the discriminative models, although (in some places) the extracted signals may be heavily mixed with the wrong speaker, given the overall SI-SDR score and the standard deviation.
Hence, the generative DiffSpEx method is preferably applied to settings where target speakers need to be extracted with high speech quality and naturalness.

\begin{figure}[t]
\centering
  \includegraphics[width=8cm]{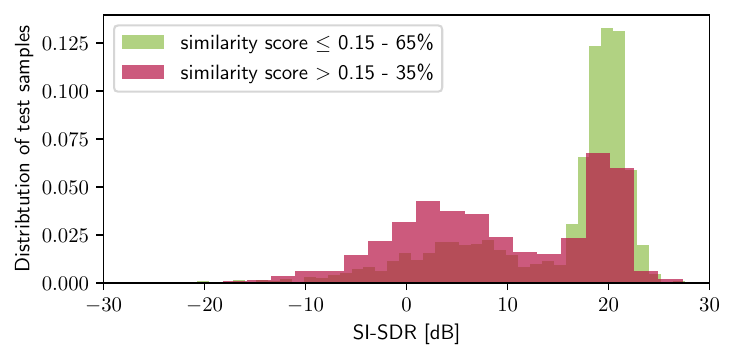}
\small\caption{SI-SDR distribution conditioned on an ECAPA-TDNN cosine similarity of $>0.15$ and $\leq0.15$ between the two source signals in the WSJ0-2mix test samples.}
\label{fig:similarity-distribution}
\end{figure}

\begin{figure}[t]
\centering
  \includegraphics[width=8cm]{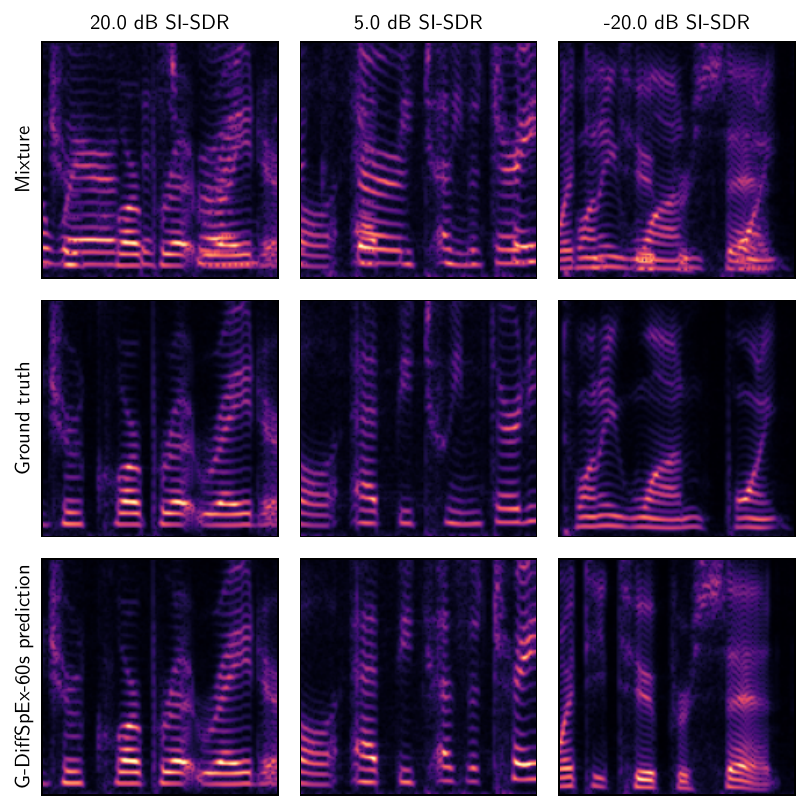}
\small\caption{Chosen evaluation samples of G-DiffSpEx-60s. In the left column, G-DiffSpEx-60s extracts the correct speaker. In the middle column, the model extracts the correct speaker in the first half of the signal segment and the wrong speaker in the second half. In the right column, the model extracts the wrong speaker. The extracted samples show natural-looking in-distribution spectrograms without distortions.}
\label{fig:overview-sample}
\end{figure}

\begin{table}[htbp]
\tabcolsep=0.05cm
\small
	\caption{Evaluation results of DiffSpEx and benchmarks on WSJ0-2mix 8~kHz \textit{min}-cut. The standard deviation is given after $\scriptstyle\pm$.} 
	\centering
	\begin{tabular}{*5c}
		\toprule
        Model &SI-SDR$\,\uparrow$&SI-SDRi$\,\uparrow$&PESQ$\,\uparrow$&NISQA$\,\uparrow$\\
		\midrule
        SpeakerBeam \cite{Zmolikova-2019, Ge-2020}&9.2&-&-&-\\
        SpeakerBeam\tablefootnote{Results generated from checkpoint provided by the \href{https://github.com/BUTSpeechFIT/speakerbeam}{SpeakerBeam} repository. Caveat: the checkpoint was trained on the bigger Libri2Mix dataset.}&$14.0\scriptstyle\pm7.3$&$11.5\scriptstyle\pm7.3$&$3.05\scriptstyle\pm0.63$&$3.27\scriptstyle\pm0.47$\\
        SpEx \cite{Xu-2020}&14.6&-&3.14&-\\
        SpEx+ \cite{Ge-2020}&18.2&-&3.49&-\\
        SpEx+\tablefootnote{Results generated from a checkpoint in the \href{https://github.com/gemengtju/SpEx_Plus}{SpEx+} repository.}&$11.6\scriptstyle\pm12.8$&$11.6\scriptstyle\pm13.2$&$3.01\scriptstyle\pm0.97$&$3.03\scriptstyle\pm0.50$\\
        SpEx-CA \cite{Wang-2021}&-&18.8&-&-\\
        X-Sepformer \cite{Liu-2023}&-&19.1&3.75&-\\
        \midrule
        G-DiffSpEx
        &$12.7\scriptstyle\pm9.3$&$10.1\scriptstyle\pm9.0$&$3.05\scriptstyle\pm1.02$&$3.55\scriptstyle\pm0.32$\\
        G-DiffSpEx-60s
        &$12.9\scriptstyle\pm9.1$&$10.3\scriptstyle\pm8.9$&$3.08\scriptstyle\pm1.00$&$3.56\scriptstyle\pm0.33$\\
		\bottomrule
	\end{tabular}
	\label{tab:TSE-results}
\end{table}
\noindent
Fig. \ref{fig:similarity-distribution} plots the distribution of the SI-SDR results for G-DiffSpEx-60s.
It reveals a bimodal distribution in performance and shows that DiffSpEx struggles with the TSE task when the speech sources in the mixture have a high ECAPA-TDNN cosine similarity, as for similarity scores $>0.15$, the distribution of SI-SDR shows a high density between -5~dB and 10~dB.
Intuitively, target speakers are harder to extract from a mixture of similar speakers.
This suggests that the model may improve, given more accurate and better matching speaker embeddings and providing more context, e.g. a bigger signal window, to the model.
\noindent
Fig. \ref{fig:overview-sample} gives chosen samples from the TSE test evaluation.

\subsection{Personalised target speaker extraction}
The evaluation results on the fine-tuned P-DiffSpEx model with different target speaker reference audio lengths are given in Table \ref{tab:P-TSE-results}, where the G-40f-DiffSpEx-60s model corresponds to a non-personalised model for speaker 40f and G-10S-DiffSpEx-600s corresponds to the same experiment repeated for 10 speakers. The postfixes to G-DiffSpEx and P-DiffSpEx indicate the total duration of reference audio in the respective personalised datasets.
\begin{table}[htbp]
\tabcolsep=0.05cm
\small
	\caption{Fine-tuning results for 40f and the set of 10 speakers (10S).} 
	\centering
	\begin{tabular}{*5c}
		\toprule
        Model &SI-SDR$\,\uparrow$&SI-SDRi$\,\uparrow$&PESQ$\,\uparrow$&NISQA$\,\uparrow$\\
		\midrule
        G-40f-DiffSpEx-60s&$14.8\scriptstyle\pm8.9$&$12.3\scriptstyle\pm8.5$&$3.43\scriptstyle\pm0.89$&$3.67\scriptstyle\pm0.18$\\
        P-40f-DiffSpEx-60s&$15.2\scriptstyle\pm7.4$&$12.6\scriptstyle\pm7.0$&$3.44\scriptstyle\pm0.77$&$3.79\scriptstyle\pm0.17$\\
        P-40f-DiffSpEx-180s&$16.2\scriptstyle\pm7.2$&$13.7\scriptstyle\pm6.8$&$3.58\scriptstyle\pm0.73$&$3.74\scriptstyle\pm0.16$\\
        P-40f-DiffSpEx-300s&$16.7\scriptstyle\pm7.0$&$14.2\scriptstyle\pm6.6$&$3.64\scriptstyle\pm0.70$&$3.74\scriptstyle\pm0.16$\\
        P-40f-DiffSpEx-600s&$17.2\scriptstyle\pm7.1$&$14.7\scriptstyle\pm6.7$&$3.69\scriptstyle\pm0.70$&$3.70\scriptstyle\pm0.16$\\
        \midrule
		G-10S-DiffSpEx-600s&$13.6\scriptstyle\pm8.5$&$10.9\scriptstyle\pm8.1$&$3.29\scriptstyle\pm0.93$&$3.34\scriptstyle\pm0.30$\\
        P-10S-DiffSpEx-600s&$16.1\scriptstyle\pm7.1$&$13.3\scriptstyle\pm6.7$&$3.60\scriptstyle\pm0.75$&$3.34\scriptstyle\pm0.30$\\
  \bottomrule
	\end{tabular}
	\label{tab:P-TSE-results}
\end{table}
\noindent
The personalised TSE models achieve a higher SI-SDR than the G-DiffSpEx model without fine-tuning.
Furthermore,  all metrics except for NISQA improve with the length of reference audio provided during fine-tuning.
This shows that DiffSpEx can be applied to settings where personalisation for particular target speakers is possible.
\section{Conclusion}
\label{sec:conclusion}
We proposed DiffSpEx, a target speaker extraction model based on score-based generative modelling and ECAPA-TDNN speaker embeddings.
We demonstrate that DiffSpEx achieves target speaker extraction performance that is better than the earlier discriminative target speaker extraction models (12.9~dB SI-SDR) while producing natural-sounding outputs per the NISQA metric (3.56).
Furthermore, we show that fine-tuning a pre-trained DiffSpEx model on one speaker in a personalised setting is possible and improves the extraction performance with as little as \SI{60}{\second} of reference audio, and more reference audio further improves the TSE performance of a personalised DiffSpEx model.
In conclusion, DiffSpEx performs well in settings that require target speaker extraction with high quality and naturalness and allow for the personalisation of target speakers.
\ninept
\bibliographystyle{IEEEbib}
\clearpage
\bibliography{refs}
\end{document}